\documentclass[aps,prb,twocolumn,showpacs,superscriptaddress,amsmath,amssymb]
{revtex4-2}
\usepackage{epsfig}
\usepackage{dcolumn}
\usepackage{graphicx}
\usepackage{amsmath}
\usepackage{amsfonts}
\usepackage{amssymb}
\usepackage{color}
\usepackage{txfonts}
\usepackage{hyperref}
\usepackage{adjustbox}
\usepackage{ulem}
\usepackage{framed}
\usepackage{wrapfig}
\usepackage{multirow}

\newcommand{\be}{\begin{equation}}
\newcommand{\ee}{\end{equation}}
\newcommand{\bea}{\begin{eqnarray}}
\newcommand{\eea}{\end{eqnarray}}

\def\k{{ {\mathbf k} }}

\def\q{{ {\mathbf q} }}
\def\Q{{ {\mathbf Q} }}

\begin{document}

\title{ Ward identity preserving approach for investigation of phonon spectrum with self-energy and vertex corrections}

\author{Sudhakar Pandey}
\affiliation{Department of Physics, North-Eastern Hill University, Shillong - 793022, Meghalaya, India}
\affiliation{Department of Physics, VIT-AP University, Amaravati - 522237, Andhra Pradesh, India}

\begin{abstract}
 
 We propose an approach for investigating the 
 many body effects on phonon spectrum in an electron-phonon coupled system by taking into account the electron self-energy and vertex corrections which respect the Ward identity  
 representing the conservation of electronic charge. Our approach provides a systematic diagrammatic expansion of the phonon self-energy $\Pi_\q(\omega)$ in powers of the  
 electron-phonon scattering strength  squared $(g^2)$.
 In this approach the many body corrections to   phonon spectrum vanish identically for the $\q = 0$ mode due to  an exact cancellation between  the contributions arising from  electron self-energy and vertex corrections. This cancellation holds irrespective of the microscopic details of the system such as dimensionality, adiabaticity, and strength of electron-phonon scattering strength etc.  Our results demonstrate that the contributions of electron self-energy and vertex corrections are not only comparable but they also tend to cancel each other so that the phonon spectrum remains nearly unaffected  due to many body effects in the regime of long-wavelength excitations ($\q \rightarrow 0$). Our results provide another constraint for the applicability of the Migdal's theorem.
 
\end{abstract}

\maketitle

\section {Introduction}
The electron-phonon coupling 
(EPC) in many materials 
is believed to give rise to a plethora of exotic quantum phenomena such as superconductivity, charge-density wave (CDW) order, polarons, etc.\cite{Giustino-RMP-2017}.  
One of the major issues related to these materials is to understand the role of many body effects arising from the EPC. 
More specifically, in systems with finite EPC, while the electron self-energy corrections are taken into account, the  vertex corrections are conventionally ignored owing to the Migdal's theorem\cite{Migdal-JETP-1958}.
However,  formulated originally for three-dimensional systems the Migdal's theorem is based on the assumption that the effective electron-phonon coupling $(\lambda)$ times  $\Omega/\epsilon_F$ is very small, i.e.,   $\lambda \Omega/\epsilon_F \ll 1$,  where $\Omega$  and $\epsilon_F$  represent  the characteristic phonon energy  and  the Fermi energy, respectively. 
Therefore, the applicability of the Migdal's theorem depends upon several constraints such as the dimensionality, adiabaticity ($\Omega/\epsilon_F$), strength of electron-phonon coupling etc.\cite{Wang-PRB-2021,Aperis-PRB-2021,Aperis-PRB-2020,Kivelson-PRB-2018,Gunnarsson-PRB-2011, Hague-JPCM-2005, Hague-JPCM-2003,Alexandrov-EPL-2001,Grimaldi-EPJ-2001, Miller-PRB-1998,Grimaldi-PRB-1995, Grimaldi-PRB-1995II, Gunnarsson-PRB-1994,Allen-Solid-State-1983}. For example, based on the phase space arguments the Migdal's theorem is speculated  to be violated  in case of one-  and two-dimensional systems \cite{Allen-Solid-State-1983}.  Similarly, the breakdown of the Migdal's theorem has been demonstrated in the strong coupling regime \cite{Alexandrov-EPL-2001}. Therefore, it is highly desirable  to formulate an approach  which can also properly address the regimes beyond the applicability of the Migdal's theorem.

Several interesting features have been demonstrated to appear due to vertex corrections both in normal and superconducting states of the electron-phonon systems \cite{Aperis-PRB-2021,Aperis-PRB-2020,Hague-JPCM-2005,Hague-JPCM-2003,Alexandrov-EPL-2001,Miller-PRB-1998,Grimaldi-PRB-1995, Grimaldi-PRB-1995II,Gunnarsson-PRB-1994}.  
Moreover, the impact of vertex corrections becomes more pronounced in the low dimensional systems\cite{Aperis-PRB-2020, Hague-JPCM-2005, Hague-JPCM-2003, Gunnarsson-PRB-1994,Allen-Solid-State-1983}. 
For example, the exact one-electron spectra in case of a one-dimensional system showing a rich satellite structure with peak separations could not be reproduced by the calculations neglecting the vertex corrections even for an effective bandwidth large compared to the phonon frequency\cite{Gunnarsson-PRB-1994}. Similarly, both the gap size and transition temperature in case of the two-dimensional system change significantly when the vertex corrections are taken into account\cite{Aperis-PRB-2020}. 

Therefore, it is necessary to do a careful analysis of the contributions of not only the electron self-energy but also  the vertex correctins for a better understanding of the many-body effects in the electron-phonon systems. Theoretically an important clue about the role of electron self-energy and vertex corrections is  provided by the Ward identity as a consequence of the conservation of electronic charge\cite{Ward-PR-1950, Mahan-Book,Schrieffer-PR-1963,Chubukov-PRB-2005}. The time-independence of the total charge requires identical vanishing of the dynamic polarizability $P_\q(\omega)$  for the $\q=0$ mode. This requirement is expressed in terms of a Ward identity which provides a specific relationship between self-energy and vertex corrections, as discussed below.

In this paper we utilize the Ward identity for formulating an approach for investigation of the many body effects on phonon spectrum in electron-phonon systems.
Our approach is based on a systematic diagrammatic expansion of the phonon self energy in powers of the electron-phonon scattering strength $(g^2)$ by taking into account the electron self-energy and vertex corrections which respect the Ward identity.
 We demonstrate that the many body corrections to the phonon spectrum 
vanish identically for the $\q = 0$ mode due to an exact cancellation between the contributions arising from electron self-energy and vertex corrections.  This cancellation suggests that the many body effects become  small in the regime of the long-wavelength excitations ($\q \rightarrow 0$)     
where the contributions from electron self-energy and vertex corrections become not only comparable but also tend to cancel each other.

\section{Model} The coupling between electrons in a single band with a branch of phonons can be described by the Hamiltonian,

\begin{eqnarray}
\hat{H}&=&\sum_\k\varepsilon_\k\;c_\k^{\dagger}c_\k+
\sum_\q\omega_\q b^\dagger_\q b_\q
\nonumber \\
& + & \sum_{\k,\q}g_\q\;c_{\k+\q}^\dagger c_\k\;(b^{\dagger}_{-\q}+b_\q)
\label{Ham}\;,
\end{eqnarray}

\noindent where $c_\k^{\dagger}$ ($c_\k$) and $b^{\dagger}_\q$ ($b_\q$) are the creation (annihilation) operators for the electron and phonon, respectively, and $\varepsilon_\k$ and $\omega_\q$ represent the bare dispersion for the electron and phonon, respectively.
Here $g_\q$ is the electron-phonon scattering strength. The momentum-dependence of the scattering strength can be considered in terms of a form factor $f_\q$ such that a vertex pair can be associated with the factor $|g_\q|^2= g^2 |f_\q|^2$ where $g$ represents the momentum-independent scattering strength. This representation for the scattering strength allows us to treat $g^2$ as an expansion parameter for the Feynman diagram series expansion\cite{Aperis-PRB-2020}.

\begin{figure}
\vspace*{-28mm}
\begin{center}
\resizebox{80mm}{50mm}{\psfig{figure=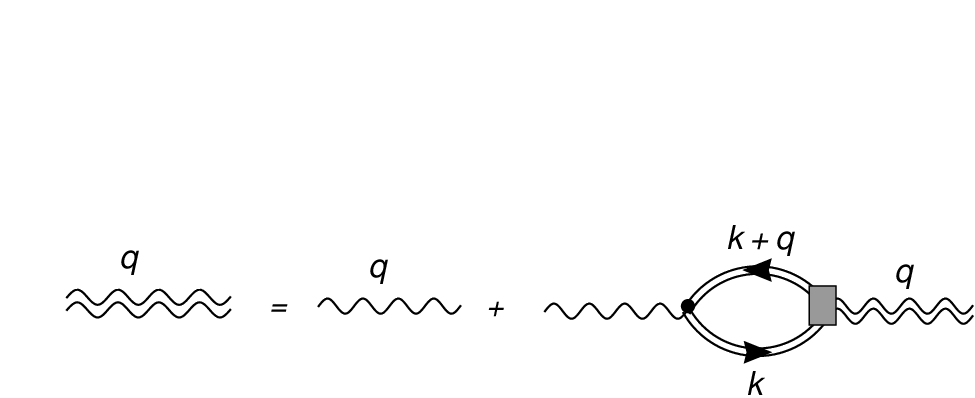}}
\end{center}
\vspace*{-5mm}
\caption{Diagrammatic representation of the Dyson's equation for the phonon Green's function. Here the single (double) wavy lines represent the bare (exact) phonon Green's function $D^0 (D)$ and the double lines with arrow represent the exact electron Green's function $G$. The shaded rectangular box represents the full electron-phonon vertex $\Gamma$.  Here we use the notations $
\q=(\q,\omega), \k=(\k,E)$}
\label{Fig-Dyson_Phonon}
\end{figure}

With an aim to study the many-body effects on phonon spectrum we consider the Dyson's equation for the phonon Green's function, as shown diagrammatically in Fig. \ref{Fig-Dyson_Phonon},

\bea
D_\q(\omega)=D_\q^0 (\omega) + D_\q^0 (\omega)\Pi_\q(\omega)D_\q(\omega)
\; ,
\eea

where $D_\q^0 (\omega)$ represents the bare  phonon Green's function
and $\Pi_\q(\omega)$
represents the phonon self-energy  
which provides an estimation of the many-body effects on phonon spectrum. 

\section{Ward identity preserving approach} This approach provides a systematic diagrammatic scheme for investigation of the many-body effects on phonon spectrum by taking into account the electron self-energy and vertex corrections, as discussed below. 
In order to study the renormalized phonon spectrum we consider the phonon self-energy $\Pi_\q(\omega)= |g_\q|^2 P_\q(\omega)$ where $P_\q(\omega)$ represents the irreducible polarizability and can be expressed as  
\bea
P_\q(\omega) &=&- i  
\sum_\k \int \frac{d E}{2\pi} G_\k(E) G_{\k+\q}(E+\omega)
\nonumber \\
& \times &
\Gamma_{\k,\k+\q}(E,E+\omega)
\label{Eq-polarizability}
\eea
Here $G$ represents the exact electron Green's function
and $\Gamma$ represents the full electron-phonon vertex. The exact electron Green's function can be expressed using the Dyson's equation $G^{-1}_\k(E)= [G^0_\k(E)]^{-1} - \Sigma_\k (E)= E -\epsilon_\k -\Sigma_\k (E)$, where  
$G^0_\k(E)$ represents the bare electron Green's function, $\epsilon_\k$ represents the bare electronic dispersion and  $\Sigma_\k(E)$ represents the  
electron self-energy owing to the electron-phonon interaction.  
We note that in general two different conventions are followed in literatures for studying the role of vertex corrections \cite{Berges-PRX-2023}. While the exact phonon self-energy involves a product of a bare and a screened electron-phonon vertex\cite{Giustino-RMP-2017}, as usually considered in the model Hamiltonian based investigations including the present paper, on the other hand, two adiabatically screened vertices are considered frequently in the first-principle based approaches\cite{Calandra-PRB-2010}.

We now utilize the Ward identity
\cite{Mahan-Book,Schrieffer-PR-1963,Chubukov-PRB-2005} 
and demonstrate that the irreducible polarizability vanishes identically for the $\q=0$ mode as follows.
 Ward identity representing the conservation of electronic charge relates the exact electron self-energy ($\Sigma$) and the
 full electron-phonon vertex ($\Gamma$) for the $\q=0$ mode as
\bea
 \Gamma_{\k,\k}(E,E+\omega)= 1+\frac{\Sigma_\k(E)-\Sigma_\k (E+\omega)}{\omega}
\eea

Using this identity the  irreducible dynamical polarizability in 
Eq.(\ref{Eq-polarizability}) for the $\q=0$ mode can be written as, 

\bea
P_0(\omega) &=& -i \sum_\k \int \frac{d E}{2\pi} G_\k(E) G_\k(E+\omega)
\Gamma_{\k,\k}(E,E+\omega)
\nonumber \\
&=&-i \sum_\k \int \frac{d E}{2\pi} \left \{ \frac {G_\k(E) - G_\k(E+\omega)}{G^{-1}_\k(E+\omega) - G^{-1}_\k(E)} \right \} 
\nonumber \\
& \times &
\left  [1+\frac{
\Sigma_\k(E)-\Sigma_\k(E+\omega)}{\omega} \right ]
\nonumber\\
&=&-i \sum_\k \int \frac{d E}{2\pi} \left \{ \frac {G_\k(E) - G_\k(E+\omega)}{\omega +\Sigma_\k(E)- \Sigma_\k(E+\omega)} \right \} 
\nonumber \\
& \times &
\left  [\frac{\omega + 
\Sigma_\k(E)-\Sigma_\k(E+\omega)}{\omega} \right ]
\nonumber\\
&=&-\frac{i}{\omega} \sum_\k \int \frac{d E}{2\pi} \left \{ {G_\k(E) - G_\k(E+\omega)} \right \}
\nonumber\\
&=& 0,
\label{Eq-q0mode}
\eea

because $\int_{-\infty}^{+\infty} d E G_\k(E)=\int_{-\infty}^{+\infty} d E G_\k(E+\omega)$, as $\omega$ gives only a constant shift in energy. 
In the second line of the above derivation we use the identity 
$AB=AB [(A-B)/(A-B)] = (A-B)/(B^{-1}- A^{-1})$.

Now we utilize this Ward identity for carrying out a perturbative expansion of phonon self-energy $\Pi_\q(\omega)$ by treating the electron-phonon scattering strength squared ( $g^2$) as an expansion parameter, 
as discussed below. We expand the electron self-energy $\Sigma$
in the Dyson's equation for electron Green's function %$G=G_0 + G_0 \Sigma G $
$G=G^0 + G^0 \Sigma G $
in powers of $g^2$ as  $\Sigma=\Sigma^{(1)}+\Sigma^{(2)}+\Sigma^{(3)}..$. Similarly, we expand the vertex corrections $\Delta \Gamma (= \Gamma-1)$
in powers of $g^2$ as 
$\Delta \Gamma=  \Gamma^{(1)}+\Gamma^{(2)}+\Gamma^{(3)} ..$.
Here $ \Gamma=1$ (i.e., $\Delta \Gamma =0$) represents the bare vertex corresponding to the case of non-interacting electrons. 
Substituting these expanded forms of $\Sigma$ and $\Gamma (=1+\Delta \Gamma$) in Eq. (\ref{Eq-polarizability}), the irreducible polarizability can also be expanded  in powers of $g^2$ as follows.

\bea
P_\q(\omega) &=& P^{(0)}_\q(\omega) + P^{(1)}_\q(\omega)+ P^{(2)}_\q(\omega)+ ...
\label{Eq-polarizability-expansion}
\eea

Here the zeroth-order term  $P^{(0)}$ represents the bare polarizability corresponding to the case of non-interacting electrons and the higher order terms represent the many body corrections to the polarizability arising due to electron self-energy and vertex corrections.

Now, as demonstrated above, since the irreducible polarizability vanishes identically for the $\q=0$ mode for an arbitrary value of $g$ as a consequence of the Ward identity, therefore, each term $P^{(n)}_\q$ (n=0,1,2,...) in 
Eq.(\ref{Eq-polarizability-expansion}) must also vanish separately order-by-order for the $\q=0$ mode. As demonstrated in the followings, it turns out that this order-by-order vanishing of the many-body corrections for the $\q=0$ mode arises due to an exact cancellation between the contributions from the electron self-energy and the vertex corrections.

We now utilize the expanded form of the irreducible polarizability in Eq. (\ref{Eq-polarizability-expansion}) for carrying out a  systematic diagrammatic expansion of the phonon self-energy %$\Pi_\q(\omega)=|g_\q^2| P_\q(\omega)$ in powers of $g^2$ 
$\Pi_\q(\omega)=|g_\q|^2 P_\q(\omega)$
as follows.
\bea
\Pi_\q(\omega)=\Pi^{(1)}_\q(\omega) + \Pi^{(2)}_\q(\omega)+\Pi^{(3)}_\q(\omega)+ ...
\label{Eq-Phonon-SE-expansion}
\eea

Here %$\Pi^{(n)}_\q(\omega)= |g_\q^2|P^{(n-1)}_\q(\omega)$
$\Pi^{(n)}_\q(\omega)= |g_\q|^2 P^{(n-1)}_\q(\omega)$ 
with n=1,2,3,...This expanded form  allows us to classify the different diagrammatic contributions to the phonon self-energy in powers of $g^2$. 

As happens in case of the irreducible polarizability, the dynamical phonon self-energy also vanishes identically for the $\q=0$ mode as a consequence of the Ward identity. Since the phonon self-energy provides a measure of the impact of the many body effects on phonon spectrum, therefore, the vanishing of the phonon self-energy for $\q=0$ mode suggests that the phonon spectrum in the regime of long-wavelength excitations ($\q\rightarrow 0$) remains nearly unaffected by the many body effects.

\begin{figure}
\vspace*{0mm}
\begin{center}
\resizebox{55mm}{35mm}{\psfig{figure=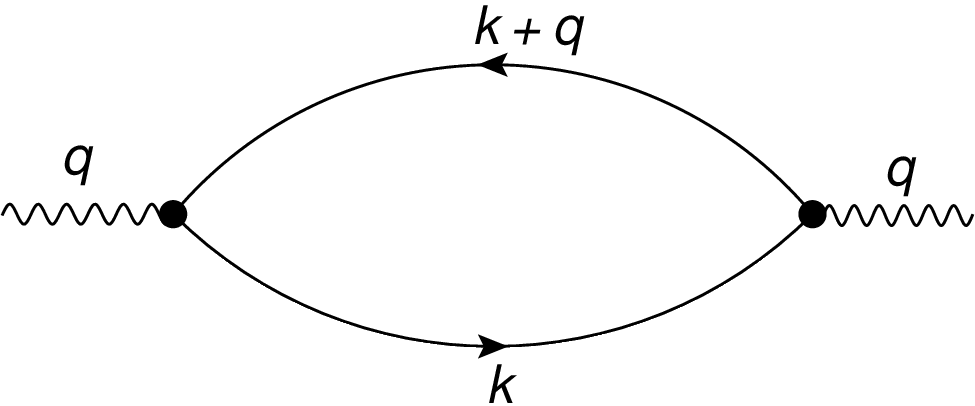}}
\end{center}
\vspace*{0mm}
\caption{Diagrammatic representation of the leading-order phonon self-energy $\Pi^{(1)}_\q(\omega)$ which involves the non-interacting electrons represented by the bare Green's function $G^0$, as shown here by single lines with arrow. Here we use the notations $\q=(\q,\omega), \k=(\k,E)$.}
\label{Fig-Pi0}
\end{figure}

\begin{figure}
\vspace*{0mm}
\begin{center}
\hspace*{-5mm}
\resizebox{100mm}{130mm}{\psfig{figure=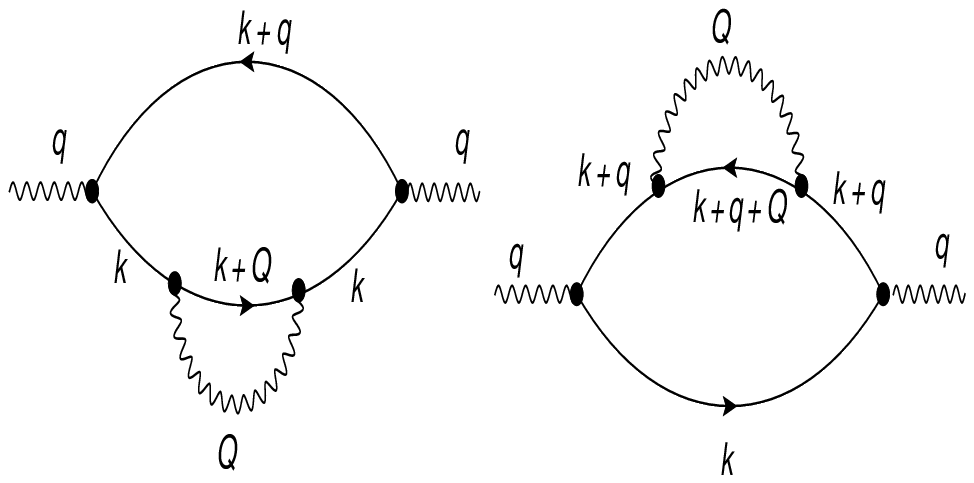}}
\end{center}
\vspace*{-85mm}
\caption{Diagrammatic representations 
of $\Pi^{(2a)}_\q(\omega)$ (left) and $\Pi^{(2b)}_\q(\omega)$ (right) which denote the leading order many-body corrections to phonon self-energy arising due to electron self-energy corrections. Here we use the notations $\q=(\q,\omega), \k=(\k,E), \Q=(\Q,\Omega)$}
\label{Fig-Pi2ab}
\end{figure}

\begin{figure}
\vspace*{-15mm}
\begin{center}
\resizebox{50mm}{40mm}{\psfig{figure=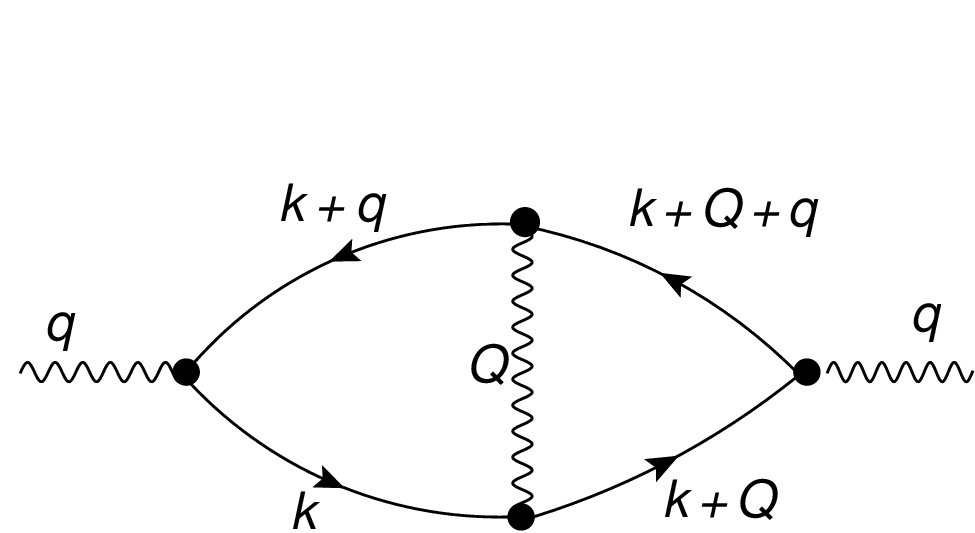}}
\end{center}
\vspace*{0mm}
\caption{Diagrammatic representation of  $\Pi^{(2c)}_\q(\omega)$ which denotes the leading order many body corrections to phonon self-energy arising due to vertex corrections. Here we use the notations $\q=(\q,\omega), \k=(\k,E), \Q=(\Q,\Omega)$}
\label{Fig-Pi2c}
\end{figure}

\section{ Many body corrections in long-wavelength limit $(\q \rightarrow 0)$} 
We now analyse the contributions of the different terms in Eq.(\ref{Eq-Phonon-SE-expansion}) in order to understand the origin of the vanishingly small many-body effects in the regime of long-wavelength excitations. For simplicity,  we consider the phonon self-energy by focusing only on the contributions of the leading and the subleading order terms in Eq. (\ref{Eq-Phonon-SE-expansion}). As shown diagrammatically in Fig. \ref{Fig-Pi0}, the leading order phonon self-energy $\Pi^{(1)}$ represents renormalization of the phonon spectrum due to  interaction between bare electrons and  phonons which can be expressed as follows.

\bea
%\Pi^{(1)}\q(\omega) &=& |g_\q|^2 P^{(0)}_\q(\omega)
\Pi^{(1)}_\q(\omega)&=& |g_\q|^2 P^{(0)}_\q(\omega)
\nonumber\\
& = &-i |g_\q|^2 \sum_k \int \frac{d E}{2\pi} G^0_\k(E)G^0_{\k+\q}(E+\omega)
\nonumber \\
\label{Eq-Pi0}
\eea

In other words, the leading order phonon self-energy $\Pi^{(1)}$ does not incorporate the many-body effects arising due to the electron self-energy and vertex corrections. These many-body effects are taken into account by the higher order terms in the expansion of the phonon self energy in Eq. (\ref{Eq-Phonon-SE-expansion}). 
We  now consider  the subleading order phonon self-energy %$\Pi^{(2)}(\q,\omega)$ 
$\Pi^{(2)}_\q(\omega)$
which can be expressed as follows.  

\bea
%\Pi^{(2)}(\q,\omega) 
\Pi^{(2)}_\q(\omega)
&=& |g_\q|^2 P^{(1)}_\q(\omega)
\nonumber\\
&=&
\Pi^{(2a)} + \Pi^{(2b)} + \Pi^{(2c)} 
\eea

Here $\Pi^{(2a)}$ and $\Pi^{(2b)}$ represent the contributions arising due to the electron self-energy  corrections, as shown diagrammatically in Fig. (\ref{Fig-Pi2ab}), and $\Pi^{(2c)}$ represents the contribution arising due to vertex corrections, as shown diagrammatically in Fig. (\ref{Fig-Pi2c}). These contributions can be expressed as follows.

\bea
\Pi^{(2a)}_\q(\omega) 
&=&-i %|g_\q^2| 
|g_\q|^2 \sum_k \int \frac{d E}{2\pi}  [G^0_\k(E)]^2 \Sigma^{(1)}_\k(E)
\nonumber \\
& \times & G^0_{\k+\q}(E+\omega) 
\label{Eq_pi2a}
\eea

\bea
\Pi^{(2b)}_\q(\omega) 
&=&-i |g_\q|^2 \sum_k \int \frac{d E}{2\pi}G^0_\k(E)
\nonumber \\
& \times &
 [G^0_{\k+\q}(E+\omega)]^2 \Sigma^{(1)}_{\k+\q}(E+\omega) 
\nonumber \\
&=& \Pi^{(2a)}_{-\q}(-\omega)
\label{Eq_pi2b}
\eea

\bea
\Pi^{(2c)}_\q(\omega) 
&=&-i |g_\q|^2 \sum_k \int \frac{d E}{2\pi} 
G^0_\k(E)G^0_{\k+\q}(E+\omega)
\nonumber \\
& \times &
\Gamma^{(1)}_{\k,\k+\q}(E,E+\omega)  
\label{Eq_pi2c}
\eea

Here $\Sigma^{(1)}$ and $\Gamma^{(1)}$ represent the  leading-order electron self-energy and vertex corrections, respectively, which  can be expressed as,

\bea
\Sigma^{(1)}_\k(E) &=& i 
\sum_\Q |g_\Q|^2
\int \frac {d\Omega}{2\pi}  G^0_{\k+\Q}(E+\Omega)
D^0_\Q(\Omega)
\; ,  
\label{Eq-electron-SE}
\eea

and

\bea
\Gamma^{(1)}_{\k,\k+\q}(E,E+\omega) 
&=& i 
\sum_\Q|g_\Q|^2 \int \frac{d \Omega}{2\pi} G^0_{\k+\Q}(E+\Omega) 
\nonumber \\
& \times & 
G^0_{\k+\Q+\q}(E+\Omega+\omega) 
D^0_\Q(\Omega)
\nonumber\\
\eea

In the followings we prove that $\Sigma^{(1)}$ and $\Gamma^{(1)}$ respect the Ward identity for the $\q=0$ mode, viz.,

\bea
 \Gamma^{(1)}_{\k,\k}(E,E+\omega)&=& \frac{\Sigma^{(1)}_\k (E) - \Sigma^{(1)}_\k(E+\omega)}
{\omega}
\label {Eq-Ward-leading-order}
 \eea

In order to prove this we consider the above expressions 
for $\Sigma^{(1)}$ and 
$\Gamma^{(1)}$ for $\q=0$ which gives

\bea
\Gamma^{(1)}_{\k,\k}(E,E+\omega) 
&=& i   \sum_\Q  |g_\Q|^2 \int \frac{d \Omega}{2\pi}
D^0_\Q(\Omega)
\nonumber \\
& \times & 
\left \{ G^0_{\k+\Q}(E')
G^0_{\k+\Q}(E'+\omega) \right \}
\nonumber \\
&=& i  \sum_\Q  |g_\Q|^2 \int \frac{d \Omega}{2\pi} D^0_\Q(\Omega)
\nonumber \\
& \times & 
\left \{ \frac{G^0_{\k+\Q}(E') - G^0_{\k+\Q}(E'+\omega)}{[G^0_{\k+\Q}(E'+\omega)]^{-1} -[G^0_{\k+\Q}(E')]^{-1}} \right \} 
\nonumber \\
&=& i  \sum_\Q   |g_\Q|^2 \int \frac{d \Omega}{2\pi}D^0_\Q(\Omega)
\nonumber \\
& \times & 
\left \{ \frac{ G^0_{\k+\Q}(E') - G^0_{\k+\Q}(E'+\omega)}{\omega} \right \} 
\nonumber \\
&=& 
\frac{\Sigma^{(1)}_\k(E) - \Sigma^{(1)}_\k(E+\omega)}
{\omega}
\eea
Here we have used the notation  %s $\k'=\k+\Q$ and 
$E'=E+\Omega$. 
  
Now, as discussed above, the net 
dynamic phonon self energy must vanish identically for the $\q=0$ mode as a consequence of the Ward identity. This condition requires that each term in the expansion of the phonon self energy must vanish separately order-by-order for the $\q=0$ mode.  

For the leading-order phonon self-energy 
$\Pi^{(1)}$ it can be demonstrated straightforwardly that $\Pi^{(1)}_{\q=0}(\omega)=0$ by substituting $\q=0$ in Eq. (\ref{Eq-Pi0}) as follows. 
\bea
\Pi^{(1)}_0(\omega) &=&  -i |g_0|^2 \sum_k \int \frac{d E}{2\pi} G^0_\k(E)G^0_\k(E+\omega)
\nonumber \\
& = &
-i |g_0|^2  \sum_k \int \frac{d E}{2\pi} 
\frac {G^0_\k(E) - G^0_\k(E+\omega)}{[G^0_\k(E+\omega)]^{-1} - [G^0_\k(E)]^{-1}}
\nonumber \\
& = &
-i \frac{|g_0|^2}{\omega} \sum_\k \int \frac{d E}{2\pi} \left \{G^0_\k(E) - G^0_\k(E+\omega)\right \}  %G_0(\k,E) - G_0(\k,E+\omega) \right \}
\nonumber \\
& = & 0
\eea
Next, we demonstrate that the subleading order dynamical phonon self energy $\Pi^{(2)}_\q(\omega)$, where many-body corrections are incorporated by taking into account the leading-order electron self-energy and vertex corrections, also vanishes for $\q=0$ mode as follows. Denoting the net contributions due to electron self energy corrections as $\Pi^{(2ab)}$,
we get 

\bea
\Pi^{(2ab)}_0(\omega)
&=& \Pi^{(2a)}_0(\omega) +  \Pi^{(2b)}_0(\omega)
\nonumber \\
&=&-i|g_0|^2  \sum_\k \int \frac{d E}{2\pi} \left \{ G^0_\k(E) G^0_\k(E+\omega) \right \}
\nonumber\\
& \times &
\left \{ G^0_\k(E)\Sigma^{(1)}_\k(E) + G^0_\k(E+\omega)\Sigma^{(1)}_\k(E+\omega) \right \}
\nonumber \\
&=&
-i|g_0|^2  \sum_k \int \frac{d E}{2\pi}
\left \{ \frac {G^0_\k(E) - G^0_\k(E+\omega)} {[G^0_\k(E+\omega)]^{-1} - [G^0_\k(E)]^{-1}} \right \} 
\nonumber \\
& \times &
\left \{G^0_\k(E)\Sigma^{(1)}_\k(E) + G^0_\k(E+\omega)\Sigma^{(1)}_\k(E+\omega)\right \}
\nonumber \\
&=&
-i |g_0|^2 \sum_\k \int \frac{d E}{2\pi}
\left \{ \frac {G^0_\k(E) - G^0_\k(E+\omega)}{\omega} \right \}
\nonumber \\
& \times &
\left \{ G^0_\k(E)\Sigma^{(1)}_\k(E) + G^0_\k(E+\omega)\Sigma^{(1)}_\k(E+\omega)\right \}
\nonumber \\
&=&
-i \frac{|g_0|^2}{\omega} \sum_\k \int \frac{d E}{2\pi}
\left \{ [G^0_\k(E)]^2\Sigma^{(1)}_\k(E)
\right. 
\nonumber \\
&  & \left .
-
G^0_\k(E)G^0_\k(E+\omega)\Sigma^{(1)}_\k(E) 
\right. 
\nonumber \\
&  & \left .
+ G^0_\k(E)G^0_\k(E+\omega)\Sigma^{(1)}_\k(E+\omega) 
 \right. \nonumber \\
& & \left.
-[G^0_\k(E+\omega)]^2\Sigma^{(1)}_\k(E+\omega) \right  \} 
\eea

Here % we can easily notice a cancellation between the first and last terms 
the first term is cancelled by the fourth term because  $\int_{-\infty}^{+\infty} d E  [ G^0_\k(E+\omega) ]^2\Sigma^{(1)}_\k(E+\omega) = \int _{-\infty}^{+\infty} d E  [ G^0_\k(E) ]^2\Sigma^{(1)}_\k(E)$. Therefore, we get
\bea
\Pi^{(2ab)}_0(\omega)
&=&
-i |g_0|^2  \sum_k \int \frac{d E}{2\pi}
G^0_\k(E)G^0_\k(E+\omega)
 \nonumber \\
& \times & 
\left \{ \frac{-\Sigma^{(1)}_\k(E)+\Sigma^{(1)}_\k(E+\omega)}{\omega} \right \} \nonumber \\
&=&
-i |g_0|^2  \sum_\k \int \frac{d E}{2\pi}
G^0_\k(E)G^0_\k(E+\omega)
\nonumber \\
& \times &
\left \{ -\Gamma^{(1)}_{\k,\k}(E,E+\omega) \right \} \nonumber \\
&=& 
-\Pi^{(2c)}_0(\omega)
\label{Eq_pi2ab}
\eea

Therefore, the net subleading order dynamic phonon  self-energy $\Pi^{(2)}_\q(\omega) [=\Pi^{(2a)}_\q(\omega) +\Pi^{(2b)}_\q(\omega) +\Pi^{(2c)}_\q(\omega)]$ vanishes identically for the $\q=0$ mode.
In other words, the contributions to phonon self energy arising due to dressing of electrons is canceled exactly from those arising due to the vertex corrections for the $\q=0$ mode. Furthermore, as obvious from our analysis,  the exact cancellation is independent of the microscopic details of the system such as dimensionality, electron density, band dispersion, phonon energy etc.

\begin{figure}
\vspace*{-10mm}
\begin{center}
\hspace*{-17mm}
\resizebox{122mm}{122mm}{\psfig{figure=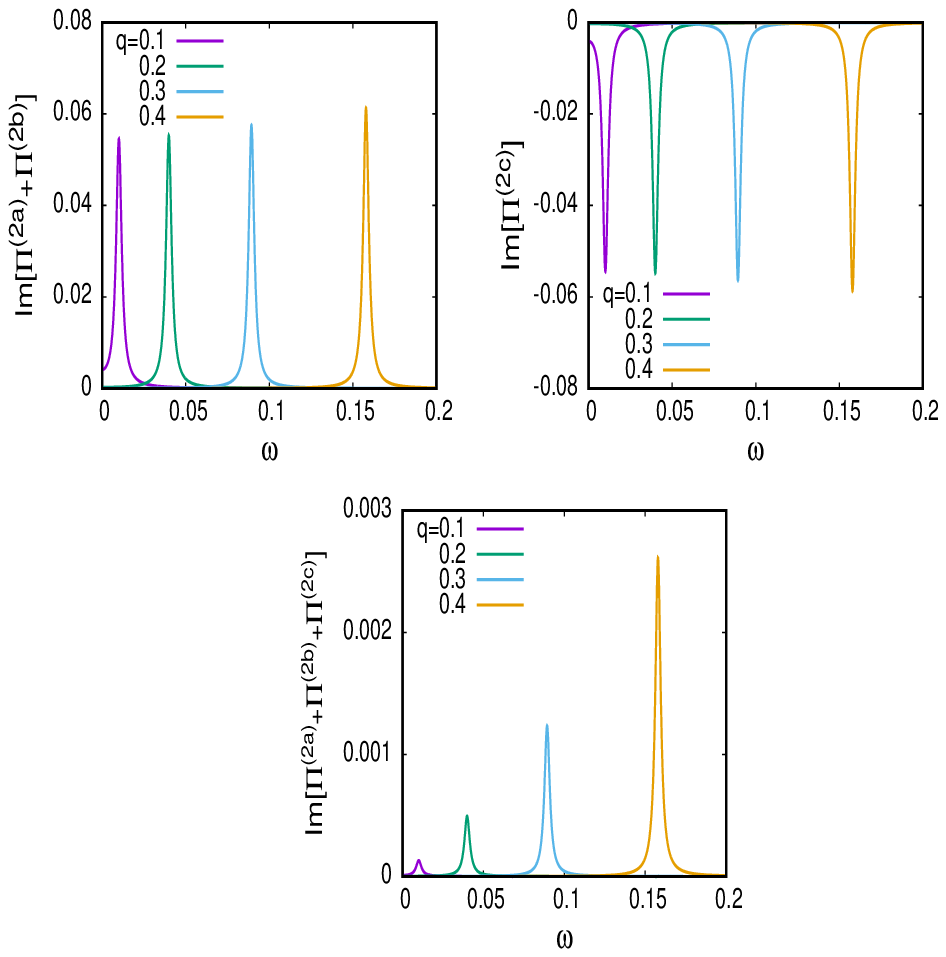}}
\end{center}
\vspace*{-40mm}
\caption{Momentum ($\q-$) dependent evolution of the contributions from the leading-order electron self-energy (upper, left) and vertex corrections (upper, right ) to the imaginary part of the phonon self-energy (lower) as shown for four different values of $\q (=0.1,0.2,0.3,0.4)$.  These results have been obtained for the  case of a single charge carrier in the one-dimensional Holstein model %  (i.e., $g_\q=g, \omega_\q= \omega$) % with band dispersion $\epsilon_\k = 2t(1-\cos\k)$ 
for parameters $\omega_0= g=t=1$  % $\omega= g=t=1$. 
As can be expected from the exact cancellation for the $\q=0$ mode, the continuous evolution of the competing contributions from the electron self-energy and vertex corrections results in a near cancellation in the long-wavelength limit $(\q\rightarrow 0)$, as can be seen for $\q=0.1$,  thereby making the net many-body correction weak.} 
\label{Fig-Pi2abc}
\end{figure}

Here it must be emphasized that while the Ward identity implies that the irreducible polarizability and hence the  phonon self-energy must vanish for the $\q=0$, however, it  does not provide a microscopic understanding of how the net many body corrections to phonon spectrum vanish for the $\q=0$ mode. In particular, it is not obvious from the Ward identity whether the net many-body correction for the $\q=0$ mode vanishes because the contributions from the electron self-energy and vertex corrections themselves vanish separately, or, it is their net sum which vanishes. The present work resolves this ambiguity and provides a clear microscopic understanding of the vanishing of the many-body corrections for the $\q=0$ mode by demonstrating  explicitly that while the contributions from both the electron self-energy and vertex corrections remain finite, the net many-body corrections vanish identically due to an exact cancellation between these two contributions as a consequence of the Ward identity. 

From the exact cancellation %for the $\q=0$ mode 
we can also naively expect a small many body correction to the phonon spectrum in the long-wavelength limit ($\q\rightarrow0$).  This is because going away from the $\q=0$ case to the $\q\rightarrow0$ limit we can expect a continuous evolution of the competing contributions from the electron self-energy and vertex corrections which will therefore tend to cancel each other because of their nearly equal and opposite contributions.
% Our prediction for  the small many body effects in the long-wavelength regimes is also supported by the quantitaive results, as shown in Fig. \ref{Fig-Pi2abc}. 
Furthermore, like the exact cancellation for the $\q=0$ mode we can also expect that the near cancellation between the contributions of the electron self-energy and vertex corrections in the long-wavelength limit will also be independent of the microscopic details of the system such as dimensionality, electron density, band dispersion, phonon energy etc. Therefore, the many-body correction to the phonon spectrum will always be small in the long-wavelength limit irrespective of the microscopic details of the system.

Our prediction for  the small many body effects in the long-wavelength regimes is also supported by the quantitative results
which can be obtained by evaluating  the subleading-order phonon self energy  $\Pi^{(2)}_\q(\omega)$ % $\Pi^{(2)}(\q,\omega)$ 
 by substituting $G^0_\k(E)=\frac{1-n_\k}{E-\epsilon_\k + i\eta} +  \frac{n_k}{E-\epsilon_\k - i\eta}$, where $n_\k=\theta(\epsilon_\k - \epsilon_F)$, and  $D^{0}_\q (\omega)= \frac{1}{\omega-\omega_\q + i\eta} - \frac{1}{\omega+\omega_\q - i\eta}$. 
For demonstrating the quantitative results  we have considered the simpler case of the one-dimensional Holstein model
% by considering
which involves a momentum-independent electron-phonon coupling ($g_\q=g$) and dispersionless bare phonon spectrum   ($\omega_\q = \omega_0$) % ($\omega_\q = \omega$) 
while the electronic dispersion %has been 
is considered within a nearest-neighbor tight-binding model  $\epsilon_\k = 2t(1-\cos\k)$, where $t$ represents the  hopping parameter. %nearest-neighbor hopping. 
In order to further simplify our analysis we consider the case of a single charge carrier, i.e., $n_\k=\delta_{\k,0}$, which interaction with the surrounding lattice distortions (phonons)  results in formation of the % well known 
quasi-particle called as polaron. Here we focus on the dominant $O(1/{\cal N})$ contributions, where ${\cal N}$ represents the total number of lattice sites. 
Fig. \ref{Fig-Pi2abc} shows the results for the momentum dependent evolution of the contributions of the electron self-energy and vertex corrections. 
As expected, while the two contributions remain finite, they almost cancel each other in the long-wavelength regimes $(\q \rightarrow 0)$, as can be seen for $\q=0.1$.
Here these results have been presented only as a representative for the purpose of quantitative demonstration
of the small many-body effects in the long-wavelength regimes and a detailed analysis will be presented elsewhere.

Although here we have demonstrated the exact cancellation of the many body corrections by focusing only on the subleading order phonon self energy $\Pi^{(2)}_\q(\omega)$, we emphasize that this cancellation will also hold order-by-order for all the higher order terms in the expansion of the phonon self-energy in Eq.(\ref{Eq-Phonon-SE-expansion}). This is because the net phonon self-energy vanishes identically for the $\q=0$ mode for an arbitrary value of $g$ as a consequence of the Ward identity which requires  the identical vanishing of the phonon self-energy for the $\q=0$ mode order-by-order in the perturbative expansion. Therefore the many body corrections to the phonon dispersion arising due to the electron self-energy must be canceled exactly by those arising due to the vertex corrections for the $\q=0$  mode at each order provided that the Ward identity is respected at each order, as demonstrated above for the subleading order phonon self-energy.

Although our present analysis for the many-body effects is focused only in the regime of the long-wavelength excitations where the charge conserving Ward identity can be utilized to understand the microscopic origin of small many body effects in terms of the competing contributions from the electron self-energy and vertex correction,  we  can also apply this approach to investigate the  momentum  dependent evolution of the two contributions 
by quantitatively evaluating the different diagrammatic contributions for an arbitrary $\q$-mode, as shown in Fig. \ref{Fig-Pi2abc}. Furthermore, since the net many body correction to phonon spectrum becomes small in the regime of long-wavelength excitations  irrespective of the microscopic details of the system, therefore, it will also be interesting to study  effects of the various parameters of the system such as dimensionality, adiabaticity, scattering strength etc., as we go away from the regime of long-wavelength excitations.  For example, as shown in Fig. \ref{Fig-Pi2abc}, in contrast to the the long-wavelength regimes, the contributions from the electron self-energy increases more rapidly in comparison to that of the vertex corrections as  we go away from the long-wavelength regimes, as can be seen for $\q=0.4$  thereby resulting in a significant net many body correction to the phonon spectrum. A detailed analysis of the evolution of the contributions from electron self-energy and vertex corrections by taking into account the various features of the system such as dimensionality, lattice structure, band filling, phonon energy, etc.  is currently under progress  will be presented elsewhere.

We note that while our analysis of the many body effects in the long-wavelength regimes is equally relevant for the different branches of phonon spectrum, it is of fundamental importance  for the acoustic phonons which possess the characteristic gapless spectrum in the regime of long-wavelength excitations with $\q=0$ mode being massless. Naively, the massless nature
can be expected from the fact that the  
 $\q=0$ acoustic phonon 
corresponds to a uniform translation of all the lattice points in a crystal which does not cost energy because it leaves the crystal as a whole invariant owing to the inherent translational symmetry. However, this seemingly simple picture becomes a highly non-trivial issue when one investigates the impact of many body effects on phonon spectrum. As evident from our analysis, taking into account only the contributions of the electron self-energy and ignoring the contribution of vertex corrections will result in a spurious finite phonon self-energy for the $\q=0$ mode, as indeed reported in literatures\cite{Loos-JPCM-2006}. 
Therefore, ignoring the contributions of the vertex corrections will not only provide a misleading estimation of the impact of the many body effects on the phonon spectrum but also destroy the characteristic gapless spectrum of acoustic phonon with a fundamentally incorrect massive $\q=0$ mode.

Apart from addressing the fundamental issues related to the many body effects on phonon spectrum, our analysis is also relevant in context of the crucial role of the long-wavelength phonons in governing the thermal properties of the low-dimensional materials. At low temperature the major contributions arise from the gapless spectrum associated with the long-wavelength acoustic phonons  because the remaining parts of the phonon spectrum with higher excitation energies are effectively frozen out.  More importantly, in case of the one-dimensional systems, where the phonon density of states $N(\omega)$ remains still finite as the phonon energy vanishes $(\omega \rightarrow 0)$, the long-wavelength acoustic phonons are predicted to give rise to anomalous thermal conductivity which increases continuously with length of the system\cite{Karamitaheri-JAP-2014}.  Indeed, the increase in thermal conductivity with increasing the length of system has been observed for several quasi-one-dimensional materials such as carbon nanotubes and silicon nanowires which are also the potential candidates for heat management and thermoelectric applications \cite{ Zhang-PhysRep-2020}. Therefore,  our analysis of the many-body effects in the long-wavelength regime is of  specific 
relevance for the  quasi one-dimensional materials.

\section {Conclusions} We have proposed  
 an approach for studying the impact of the many-body effects on the phonon spectrum  due to electron-phonon coupling. This approach is based on a systematic diagrammatic expansion of the phonon self-energy which allows to 
incorporate the many-body corrections in terms of the electron self-energy and vertex corrections which respect the Ward identity representing the conservation of electronic charge.  We have demonstrated that irrespective of the microscopic details of the system such as dimensionality, adiabaticity, and strength of electron-phonon coupling etc., the net many body corrections to phonon spectrum
 vanish identically for the $\q=0$ mode due to an exact cancellation between the contributions arising from electron self-energy and vertex corrections. 
  Our analysis predicts that the phonon spectrum remains nearly unaffected by the many-body effects in the regime of the long-wavelength excitations ($\q \rightarrow 0$) where
  the contributions of electron self-energy and vertex corrections become not only comparable but they also  tend to cancel each other. Our results provides another constraint for the applicability of the Migdal's theorem, viz., the vertex corrections can not be neglected in the regime of the long-wavelength excitations.
A detailed quantitative investigation based on this approach  is currently  under progress and will be presented elsewhere.

\section{ Acknowledgments} This work is a part of a broader collaboration with O. S. Bari\v si\' c and J. Krsnik, supported by the Croatian Science Foundation Project No. IP-2016-06-7258

\end{document}